\documentclass[nohyper,12pt,a4paper]{JHEP}
\usepackage{epsfig}

\title{Taking the Temperature of a Black Hole}

\author{Erling J. Brynjolfsson$^{1),2)}$ and Larus Thorlacius$^{1),2)}$\\ 
1) University of Iceland, Science Institute \\
Dunhaga 3, 107 Reykjavik, Iceland\\
\ \\
2) NORDITA, Roslagstullsbacken 23 \\
106 91 Stockholm, Sweden \\ 
\ \\
E-mail: \email{erlingbr@hi.is}, \email{lth@hi.is}}

\abstract{We use the global embedding of a black hole spacetime into a higher dimensional
flat spacetime to define a local temperature for observers in free fall outside a static black hole. 
The local free-fall temperature remains finite at the event horizon and in asymptotically flat
spacetime it approaches the Hawking temperature at spatial infinity. Freely falling observers 
outside an AdS black hole do not see any high-temperature thermal radiation even if the 
Hawking temperature of such black holes can be arbitrarily high.}


\preprint{NORDITA-2008-16, RH-04-2008}

\begin{document}


\section{Introduction}
A black hole in asymptotically flat spacetime emits thermal radiation 
with characteristic temperature
\begin{equation}
T_H = \frac{\kappa}{2\pi} ,
\label{thawking}
\end{equation}
where $\kappa$ is the surface gravity of the black hole \cite{Hawking:1974sw}.
The Hawking temperature $T_H$ is the temperature of the radiation as measured by 
asymptotic observers. The local temperature measured at a finite distance from the
black hole will in general be different from $T_H$ and depends
on the state of motion of the observer carrying out the measurement. 
For example, the temperature measured by a fiducial observer, {\it i.e.} an observer 
who remains at rest with respect to the black hole at a fixed distance, will differ 
from the temperature measured by an observer in free fall at the same distance.

For a spherically symmetric static black hole, described by a line
element of the form
\begin{equation}
ds^2 = -f(r)dt^2 + \frac{1}{f(r)}dr^2 + r^2d\Omega^2 ,
\label{metric}
\end{equation}
the local fiducial temperature is given by 
\begin{equation}
T_{FID}(r) = \frac{T_H}{\sqrt{f(r)}} .
\label{tfid}
\end{equation}
The fiducial temperature $T_{FID}$ diverges at the black hole event horizon and
in asymptotically flat spacetime $T_{FID}$ approaches the Hawking temperature
asymptotically far away from the black hole. In asymptotically anti de Sitter spacetime,
on the other hand, the fiducial temperature goes to zero far away from the black hole.

There are several ways to obtain $T_{FID}$. 
It can, for example, be inferred from the transition rate of a particle detector interacting 
with the radiation field at a fixed position with respect to the black hole, as was shown 
by Unruh \cite{Unruh:1976db}. In the same paper, Unruh also showed that a uniformly 
accelerated observer in flat spacetime, with proper acceleration $a$, will detect 
thermal radiation at the so-called Unruh temperature,
\begin{equation}
T_U = \frac{a}{2\pi} .
\end{equation}
The Hawking effect and the Unruh effect are related. Both arise from a mismatch
between vacuum states employed by different observers and both involve regions of 
spacetime that are hidden behind horizons.\footnote{The two effects are not equivalent, 
however. In particular, the quantum states that enter into the two calculations are 
different (for a review see \cite{Wald:1999vt}).} 
The connection between the Unruh effect and the Hawking effect is particularly 
apparent for a large Schwarzschild black hole, for which the near-horizon region is 
almost flat. In this case, a fiducial observer who remains fixed close to the event horizon 
measures high fiducial temperature (\ref{tfid}), but this observer is also highly accelerated 
in (almost) flat spacetime.

There is also a more indirect relation between the two effects that involves the global
embedding of the black hole spacetime into a higher dimensional flat spacetime
\cite{DLads,Deser:1998xb}. In this context the Hawking effect for a fiducial observer 
in the black hole spacetime can be interpreted as the Unruh effect for a uniformly 
accelerated observer in the higher dimensional flat spacetime. This relationship has
been mapped out for a variety of static, spherically symmetric spacetime geometries
by Deser and Levin \cite{Deser:1998xb} and will be briefly reviewed 
below. The local fiducial temperature and the Unruh temperature in a flat embedding 
geometry have also been shown to match for various $d$-dimensional static black 
holes in \cite{Santos:2004ws}. 

The main goal of the present paper is, however, to obtain the local temperature as
measured by observers in free fall towards a black hole. The notion of a local free-fall
temperature is less precise than that of a fiducial temperature. This is because a
detector in free fall is not static but moves in an environment that is a function of
the radial variable $r$ and will therefore in general measure a spectrum of radiation
that is not precisely thermal. On the other hand, for a macroscopic black hole and an 
observer who is moving sufficiently slowly with respect to the black hole rest frame
the change in the environment will be slow compared to the microscopic processes
involved inside a particle detector, allowing temperature readings to be made as 
the observer progresses along the free fall orbit. Even if these temperature readings 
do not correspond to a precise thermal state they provide an effective temperature 
that indicates the amount of heating that an observer in free fall would experience along 
a given orbit. See \cite{Korsbakken:2004bv} for a related discussion involving observers
in general stationary accelerated frames in flat spacetime.

We obtain the local free-fall temperature by extending the above embedding method
to a certain class of freely falling observers. Our proposal involves two key steps. 
One is to consider the Unruh effect for accelerated observers whose proper acceleration 
is not uniform but changes with time. As long as the rate of change of the acceleration is 
slow on the timescale of temperature measurements with whatever particle detector a 
given observer is carrying we may talk about a local Unruh temperature being measured
along the observer's orbit. 

The other key step is to extend the embedding method to include a certain class of 
observers in free fall, which we refer to as 'freely falling at rest', or FFAR, observers.
As emphasized by Deser and Levin  \cite{Deser:1998xb} the matching between the local
temperature measured by a fiducial observer and the Unruh temperature of the corresponding 
uniformly accelerated observer in the higher-dimensional flat spacetime rests on the world-line 
of the fiducial observer being tangent to a time-like Killing vector in the black hole spacetime.
Or, in more physical terms, that the fiducial observer is at rest in a static geometry. The fact
that geodesic orbits are not tangent to the time-like Killing vector suggests that the embedding 
method cannot be applied to observers in free fall but we note that there are special points
where the tangent vector to a geodesic orbit is in fact parallel to the time-like Killing vector.
These are the turning points of radial geodesics where freely falling observers are 
momentarily at rest with respect to the black hole. Such a 'freely-falling observer at rest', or 
FFAR observer, can be obtained by switching off the acceleration keeping a fiducial observer 
in place. Immediately after the release the formerly fiducial observer is in free fall but has not 
yet began moving towards the black hole.  

The FFAR condition at a given spatial location outside the event horizon determines a unique
radial orbit that has its turning point at that location. The radial orbit in turn defines a curve in 
the higher dimensional flat embedding spacetime and this curve can be viewed as the 
worldline of an accelerated observer in the flat spacetime. In general the acceleration of this 
observer is not uniform but for a macroscopic black hole we will have a local Unruh 
temperature defined at the point on the curve that corresponds to the turning point of the radial 
orbit, and it is this local Unruh temperature that we identify as the local free fall temperature in
the black hole geometry. We note that it seems natural to consider FFAR observers when 
discussing the black hole temperature as measured in free fall because they are precisely
those freely falling observers who are at rest, if only momentarily, in the black hole rest frame. 
As discussed above, even FFAR observers will detect a spectrum of radiation that is only
approximately thermal. All other observers in free fall are moving with respect to the source of 
the Hawking radiation and therefore detect a spectrum of radiation that is further 'transformed'
away from thermal.

We consider a number of examples in subsequent sections of the paper to illustrate the
procedure and to check if it provides physically reasonable answers for a local free-fall 
temperature around various four-dimensional black holes including Schwarzschild, 
AdS-Schwarzschild, and Reissner-Nordstr\"om. The case of so-called large 
AdS-Schwarzschild black holes is particularly interesting since the definition (\ref{thawking}) 
gives a Hawking temperature that grows without bound with increasing black hole mass, 
but in recent work by S.~Hemming and one of us \cite{HL} it was claimed that observers in free 
fall would nevertheless measure very low ambient temperature near one of these large 
AdS black holes. This expectation is confirmed by the direct calculation of a free-fall
temperature for AdS-Schwarzschild in section~\ref{adss}. 

It would be interesting to compare our results obtained by the embedding method
to a more conventional calculation involving an detector coupled to a quantum 
field in a black hole background geometry \cite{Unruh:1976db}. 
The fact that an observer who falls into a
black hole is not stationary, and runs into the black hole singularity within a finite proper
time, makes it awkward to extract the detector transition rate per unit time from such a
calculation. There is, however, another class of freely falling observers for which a 
conventional computation of the Hawking effect can be carried out, namely geodesic
observers in stable circular orbits around a spherically symmetric black hole. 
Such a calculation was performed by Chen {\it et al.} in \cite{Chen:2004qw}. 
They also applied the embedding method to this class of observers and compared 
results obtained by the two methods. A circular orbit around a Schwarzschild black hole 
is mapped into a curve that involves rotation in addition to acceleration in the 5+1 
dimensional flat embedding spacetime. A generalized Unruh effect involving motion
along a general stationary curve of this type in 3+1 dimensional Minkowski spacetime
was studied by Korsbakken and Leinaas in \cite{Korsbakken:2004bv}. It is straightforward
to extend their results to higher dimensions and Chen {\it et al.} found a perfect match
between the local temperature measured by a geodesic observer in a circular orbit around
a Schwarzschild black hole and the generalized Unruh temperature of the corresponding 
observer in the higher-dimensional embedding space,
\begin{equation}
T_{CIRC}(r) = \frac{T_H}{\sqrt{1-3m/r}} ,
\label{tcirc}
\end{equation}
where $m$ is the black hole mass. We will not repeat their calculations here but refer
to \cite{Korsbakken:2004bv} and \cite{Chen:2004qw} for details.

In the present paper we are primarily interested in observers in radial free fall into a
black hole. As mentioned above, such observers have the drawback that they are not
undergoing stationary motion but since they eventually fall into the black hole they 
probe the near horizon region where there are no stable circular orbits.

\section{Higher-dimensional embedding}
Let $S$ be a $d$ dimensional spacetime manifold with metric $g_{\mu\nu}$ that is 
embedded into a $D$ dimensional Minkowski spacetime $M$ with a metric $\eta_{IJ}$ of 
mostly plus signature and with one or more time dimensions. Here the Greek indices run 
from 0 to $d-1$ and the capital Roman indices from 0 to $D-1$, $D > d$. Writing the 
embedding functions,
\begin{equation}
Z^I = Z^I(x^\mu)
\end{equation}
the metric of $S$ is induced from the higher-dimensional flat metric through
\begin{equation}
g_{\mu\nu} = \eta_{IJ}\frac{\partial Z^I}{\partial x^\mu}\frac{\partial Z^J}{\partial x^\mu}.
\end{equation}
Such an embedding always exists \cite{Goenner} but obtaining a global embedding in
the presence of an event horizon requires some care.

The tangent space $T(M)$ of the Minkowski spacetime, with basis 
$\{\frac{\partial}{\partial z^I}\}$, can be written $T(M) = T(S)\bigoplus N(S)$, where 
$T(S)$ is the tangent space of $S$, with 
basis $\{\frac{\partial}{\partial x^\mu}\}$, and $N(S)$ is orthogonal to $T(S)$. 
We choose a basis $\{\hat\xi_m\}$ in $N(S)$ such that it is orthonormal with respect to the 
Minkowskian metric of $M$, {\it i.e.} $\eta(\hat \xi_m,\hat \xi_n) = \eta_{mn}$. In the examples
we consider below $N(S)$ is either Euclidean or has one time-like dimension.

If $x^\mu(\tau)$ is a timelike curve in $S$ then the tangent vector $u=\frac{d}{d\tau}$ can be 
written \cite{Goenner, analysis}
\begin{equation}
\frac{du}{d\tau} = \nabla_u u + \alpha(u,u) .
\label{tangentvec}
\end{equation}
Here $\nabla_u$ denotes the covariant derivative in $S$ along $u$ and $\alpha$ is the second fundamental form of $S$. We have that $\nabla_u u \in T(S)$ and $\alpha(u,u) \in N(S)$ so that 
when we square equation (\ref{tangentvec}) we obtain the Pythagorian relation,
\begin{equation}
a_D^2 = a_d^2 + \alpha^2
\label{pythago}
\end{equation}
where 
\begin{eqnarray}
a_D^2 =  \eta_{IJ}a_D^Ia_D^J, & \qquad  & \frac{du}{d\tau} 
=  a_D^I\frac{\partial}{\partial z^I}, \\
a_d^2 = g_{\mu\nu}a_d^\mu a_d^\nu, & \qquad & \nabla_uu 
= a_d^\mu\frac{\partial}{\partial x^\mu}, \nonumber \\
\alpha^2 = \eta_{mn}\alpha^m\alpha^n, &\qquad & \alpha(u,u) 
= \alpha^m\hat\xi_m . \nonumber
\end{eqnarray}
The $a_D^I$ are the components of the acceleration vector of an observer in the 
$D$-dimensional embedding space moving along the image of the original curve while $a_d^\mu$ are the components of the acceleration in $S$.
 
In the case of a fiducial observer in a static spherically symmetric spacetime $S$ the 
corresponding $D$-dimensional observer is uniformly accelerated. If the constant
$D$-acceleration is spacelike it defines an Unruh temperature which agrees with 
the local fiducial temperature in $S$ \cite{Deser:1998xb}. If, on the other hand, the
$D$-acceleration is timelike the Unruh temperature is formally purely imaginary and
the fiducial observer in $S$ does not detect any thermal radiation. An example of 
the latter behavior is provided by fiducial observers in an otherwise empty AdS 
spacetime, {\it i.e.} observers who are accelerated so as to remain in a fixed 
position relative to a reference observer at the origin in AdS space \cite{Deser:1998xb}.

For FFAR observers we find the image under the embedding of the turning point of the 
corresponding radial orbit, evaluate the acceleration $D$-vector there, and take the 
resulting local Unruh temperature in $M$ to define a local temperature for observers
in free fall in $S$. 
We consider several different static spherically symmetric geometries of the form 
(\ref{metric}) and verify that the free-fall temperature so obtained meets a number of
criteria that can be expected on physical grounds. In particular, the free-fall temperature 
is finite at the event horizon of a black hole while the fiducial temperature is divergent. 
This follows immediately from the embedding relation (\ref{pythago}) since $a_d=0$ for 
an observer in free fall while for a fiducial observer $a_d\rightarrow\infty$ at the event 
horizon.

\section{Schwarzschild black holes}
Consider a four-dimensional Schwarzschild black hole. The line element is given by
(\ref{metric}) with 
\begin{equation}
f(r) = 1- \frac{2m}{r} ,
\end{equation}
where $m$ is the black hole mass. A global embedding of the Schwarzschild geometry
into six-dimensional Minkowski spacetime with metric 
$\eta_{IJ} = \textrm{diag}(-1,1,\ldots,1)$ was found by Fronsdal \cite{Fronsdal:1959uu},
\begin{eqnarray}
\label{SGEMS}
Z^0 &=& 4m\,\sqrt{1-2m/r}\,\sinh(t/4m), \\\nonumber
Z^1 &=& 4m\,\sqrt{1-2m/r}\,\cosh(t/4m), \\\nonumber
Z^2 &=& \int \mathrm{d} r\;\sqrt{2m/r + 4m^2/r^2 + 8m^3/r^3},\\\nonumber
Z^3 &=& r\sin\vartheta\cos\varphi, \\\nonumber
Z^4 &=& r\sin\vartheta\sin\varphi, \\\nonumber
Z^5 &=& r\cos\vartheta.
\end{eqnarray}
Now consider a freely falling observer, dropped from rest at $\tau=0$ at $r=r_0$. 
The equations for the orbit are
\begin{equation}
\frac{dt}{d\tau} = \frac{\sqrt{1-x_0}}{1-x}, \qquad\qquad 
\frac{dx}{d\tau}= \frac{x^2}{2m}\sqrt{x-x_0},
\end{equation}
where we have introduced a dimensionless radial variable
$x\equiv 2m/r$ which runs from $x=0$ at $r\rightarrow\infty$ to $x=1$ at the
event horizon. The 6-acceleration is spacelike for all timelike orbits and at 
the turning point $x=x_0$, where the observer is dropped from rest, its magnitude 
is given by
\begin{equation}
a_6 = \frac{1}{4m}\sqrt{1 + x + x^2 + x^3}.
\end{equation}
Taking the local temperature measured by a freely falling observer at rest to be the local 
Unruh temperature of the corresponding observer in the six-dimensional flat spacetime 
we obtain
\begin{equation}
T_{FFAR} = \frac{1}{8\pi m}\sqrt{ 1 + x + x^2 + x^3 }.
\end{equation}
We see that asymptotically far from the black hole $T_{FFAR}$ reduces to the Hawking 
temperature $T_H=1/8\pi m$ and then gradually rises to $T_{FFAR} = 2T_H$ as the 
horizon is approached. This is in line with physical expectations. In contrast with fiducial
observers, who detect high-temperature radiation in the region near the event horizon due
to the strong acceleration that is needed to keep them in place, observers in free fall will 
only detect a smooth rise by an order-one factor in the temperature going from the asymptotic
region towards the horizon. Figure~\ref{fig:temp_comp} plots the two temperatures 
$T_{FID}$ and $T_{FFAR}$ as a function of the dimensionless radial variable $x$. 
\begin{center}
\FIGURE{
\includegraphics{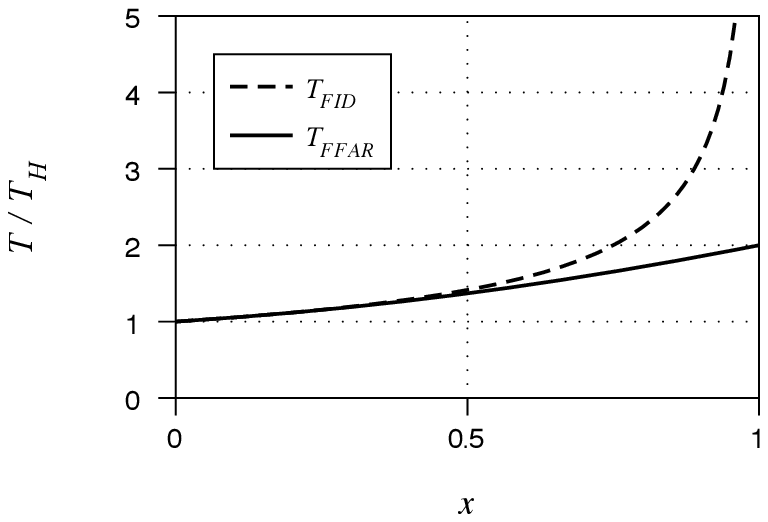}
\caption{The local temperatures $T_{FFAR}$ and $T_{FID}$ plotted as a 
function of the dimensionless radial variable $x=2m/r$. The fiducial temperature $T_{FID}$ blows 
up at the horizon while the free-fall temperature $T_{FFAR}$ remains finite.}
\label{fig:temp_comp}
}
\end{center}

We note that our method gives a physically reasonable answer for $T_{FFAR}$ at all values 
of $r\geq 2m$. However, the precise numerical value $T_{FFAR}=2T_H$ at the event horizon 
has limited operational meaning. First of all, as was discussed in the introduction, the local
free-fall temperature is not a precise notion. On top of this, a freely-falling observer passing
through the horizon has only a proper time of order $m$ left before running into the curvature 
singularity at $r=0$ and since the characteristic wavelength of thermal radiation at $T\sim 2T_H$ 
is also of order $m$ the observer cannot measure temperature to better than 
${\mathcal O}(1/m)$ precision 
near the horizon. Although our result for the free-fall temperature is thus only qualitative in the 
region near the event horizon, it does confirm the expectation expressed in early work of 
Unruh \cite{Unruh:1977ga} that an infalling observer will not run into highly energetic particles 
at the horizon. 

\section{AdS-Schwarzschild black holes}\label{adss}

Now consider a black hole in 3+1 dimensional spacetime with negative cosmological constant 
$\Lambda=-3/\ell^2$. The line element is given by (\ref{metric}) with
\begin{equation}
f(r) = 1 - \frac{2m}{r} + \frac{r^2}{\ell^2},
\end{equation}
where $m$ is the mass of the black hole. The event horizon is located at $r=r_H$, where 
$r_H$ is the real root of $1-2m/r + r^2/\ell^2$ and the surface gravity is 
\begin{equation}
\kappa=\frac{\ell^2+3r_H^2}{2r_H\ell^2}.
\end{equation}
The surface gravity has a minimum value, 
$\kappa\geq\sqrt{3}/\ell$, and each value above the minimum one is realized for two
values of $r_H$, corresponding to a large AdS black hole with $r_H>\ell/\sqrt{3}$ and a small AdS
black hole with $r_H<\ell/\sqrt{3}$. Large AdS black holes with $r_H\gg\ell$ correspond to 
high-temperature thermal states in a dual gauge theory \cite{Witten:1998zw} while
small black holes with $r_H\ll\ell$ can be viewed as more-or-less ordinary 
Schwarzschild black holes in a 'cosmological' background with a negative cosmological
constant.

The Hawking temperature (\ref{thawking}) of AdS black holes on 
the 'large' branch grows linearly with $r_H$, and becomes arbitrarily high for very large black 
holes. As we will see below, this does not mean that the physical temperature measured by an 
observer in free fall becomes large outside large AdS black holes.

The AdS-Schwarzschild geometry can be globally embedded into a 7-dimensional flat 
spacetime with the metric $\eta_{IJ} = \textrm{diag}(-1,1,\ldots,1,-1)$, which has two time-like 
dimensions \cite{Deser:1998xb}. The embedding functions are given by
\begin{eqnarray}
Z^0 &=& \kappa^{-1}\,\sqrt{f(r)}\,\sinh(\kappa t), \\\nonumber
Z^1 &=& \kappa^{-1}\,\sqrt{f(r)}\,\cosh(\kappa t), \\\nonumber
Z^2 &=& \int \mathrm{d} r\;\frac{\ell^3+\ell r_H^2}{\ell^2+3r_H^2}\sqrt{\frac{r^2r_H 
+ rr_H^2+r_H^3}{r^3(r^2+rr_H+r_H^2+\ell^2)}}, \\\nonumber
Z^6 &=& \int\mathrm{d} r\;\frac{1}{\ell^2+3r_H^2}\sqrt{\frac{(\ell^4+10\ell^2r_H^2+9r_H^4)
(r^2+rr_H+r_H^2)}{r^2+rr_H+r_H^2+\ell^2}},
\end{eqnarray}
and $Z^3, Z^4, Z^5$ are three-dimensional spherical coordinates as in equation (\ref{SGEMS}). 

Calculating the FFAR acceleration as before, one finds the following expression for the
free-fall temperature squared,
\begin{equation}
\label{adsffar}
T_{FFAR}^2 = \frac{1}{16\pi^2\ell^2}\;\frac{-4(1+x) + (c^2+1)(c^2+5)x^2 
+ (c^2+1)^2(1+x+x^2)x^3}{1+x+(c^2+1)x^2},
\end{equation}
where $x=r_H/r$ and $c\equiv\ell/r_H$. Figure~\ref{AdSSslices} plots the free-fall 
temperature squared for black holes on the two branches: A a large one with $c=0.5$ and 
a small one with $c=100$. In both cases $T_{FFAR}^2$ rises monotonically with $x$, from a 
negative value at $x=0$ to a positive value at the event horizon at $x=1$. The negative
value of $T_{FFAR}^2$ at spatial infinity reflects the fact that there is no thermal radiation
in the empty AdS region asymptotically far away from the black hole. Near the horizon, 
on the other hand, the free-fall temperature is real valued. 
\begin{center}
\FIGURE{
\includegraphics{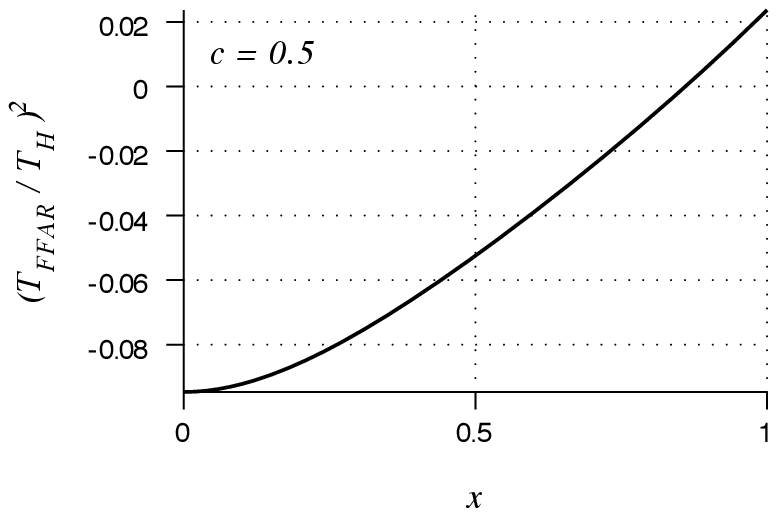}
\includegraphics{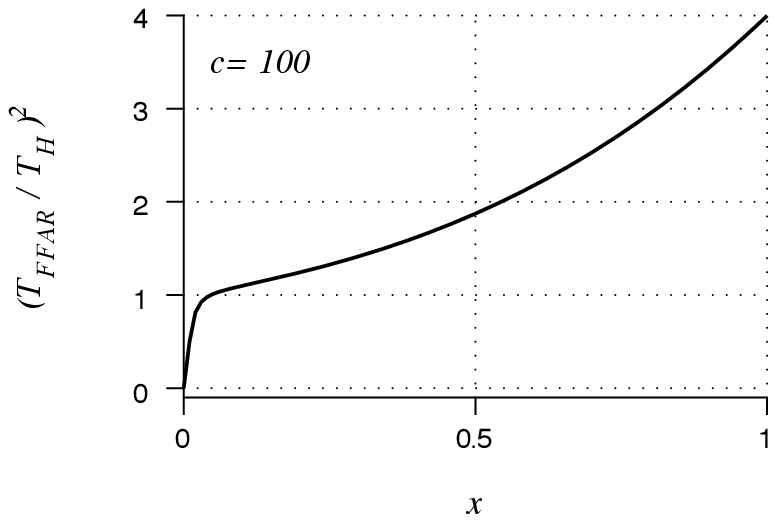}
\caption{$T_{FFAR}^2$ plotted in units of $T_H^2$ for $c=1/2$ and $c=100$. A negative value of $T_{FFAR}^2$ is interpreted as having no thermal radiation in that region.}
\label{AdSSslices}
}
\end{center}

The expression (\ref{adsffar}) for the free-fall temperature simplifies in the 
two limits $r\to\infty$ and $r\to r_H$, {\it i.e.} $x\to 0$ and $x\to 1$ respectively. 
For $r\to\infty$ one obtains
\begin{equation}
T_{FFAR}^2 \to -\frac{1}{4\pi^2 \ell^2},
\end{equation}
which is precisely the answer found for a geodesic observer in empty AdS 
space \cite{DLads}. At the event horzion, $r\to r_H$, we instead find
\begin{equation}
T_{FFAR}^2 \to\frac{1}{4\pi^2r_H^2}.
\end{equation}
For a small AdS black hole with $\ell \gg r_H$ the horizon area reduces to that of a Schwarzschild
black hole in asymptotically flat spacetime, $r_H \approx 2m$, and (\ref{adsffar}) reduces to 
$T_{FFAR}\to 2 T_H$ at the event horizon. As we move away from the black hole, but remain 
within the region $r_H<r<\ell$, the free-fall temperature approaches the Hawking temperature.
When we get to cosmological length scales $r>\ell$ the geometry approaches that of empty AdS 
spacetime and $T_{FFAR}^2$ turns negative. This behavior is evident in the lower plot in 
Figure~\ref{AdSSslices} .

On the other hand, for a large AdS black hole with $r_H\gg\ell$ we find that at the event horizon
\begin{equation}
T_{FFAR}\to\frac{1}{2\pi r_H} \ll T_H .
\end{equation}
The characteristic wavelength of the thermal radiation is then $\lambda \sim 1/T_{FFAR} \gg \ell$
and at such low temperatures one expects the radiation to be mainly in low angular modes.
The value of $r$ where the free-fall temperature squared changes sign indicates the size of 
the radial 'box' that the thermal radiation is confined to. It can be easily verified that $T_{FFAR}^2$ 
in equation (\ref{adsffar}) equals zero at the following value of the dimensionless variable $x$,
\begin{equation}
x=x_0 = -\frac{1}{2} + \sqrt{\frac{1}{4}+\frac{2}{c^2+1}}.
\end{equation}
For very large AdS black holes, we can expand in the small parameter $c= \ell/r_H \ll 1$.
The $r$-coordinate location of the zero is $r=r_0$ where
\begin{equation}
\frac{r_0}{r_H} = \frac{1}{x_0} 
= 1 + \frac{2}{3}\left(\frac{\ell}{r_H}\right)^2 + \mathcal O \left( \frac{\ell}{r_H}\right)^4,
\end{equation}
and the spatial proper distance $D_0$ from the horizon to the point $r=r_0$ is 
\begin{equation}
D_0 = \frac{2\sqrt{2}}{3}\left(\frac{\ell}{r_H}\right)\ell + \ldots .
\end{equation}
The thermal radiation is confined to a layer of thickness $D_0\ll\ell$ surrounding
the horizon of the large AdS black hole. Since the thickness of the layer is much smaller than the 
characteristic wavelength of thermal radiation at the local free-fall temperature
the radiation field outside the black hole will be in the form of an evanescent wave rather than 
propagating thermal modes. It therefore appears that an observer in free fall outside a large
AdS black hole will not see thermal radiation at all, even if the Hawking temperature of the
black hole is very high.

\section{Reissner-Nordstr\"om black holes}
Finally, we calculate the free-fall temperature for a Reissner-Nordstr\"om black hole. 
The line element is given by (\ref{metric}) with
\begin{equation}
f(r) = 1-\frac{2m}{r}+\frac{e^2}{r^2},
\end{equation}
where $m$ is the mass and $e$ is the electric charge of the black hole. The geometry 
has two horizons at $r_\pm = m\pm\sqrt{m^2-e^2}$, the zeros of $f(r)$.

It is not necessary to find a global embedding of the Reissner-Nordstr\"om spacetime into a
higher-dimensional flat spacetime in order to obtain the temperature outside the outer horizon
using the embedding method. It is sufficient to find an embedding that covers the region 
outside the inner horizon \cite{Deser:1998xb},
\begin{eqnarray}
Z^0 &=& \kappa^{-1}\,\sqrt{f(r)}\,\sinh(\kappa t), \\\nonumber
Z^1 &=& \kappa^{-1}\,\sqrt{f(r)}\,\cosh(\kappa t), \\\nonumber
Z^2 &=& \int \mathrm{d} r\;\sqrt{\frac{r^2(r_++r_-) + r_+^2(r_+ +r)}{r^2(r-r_-)}},\\\nonumber
Z^6 &=& \int\mathrm{d} r\;\sqrt{\frac{4r_+^5r_-}{r^4(r_+-r_-)^2}},
\end{eqnarray}
where the surface gravity is $\kappa = (r_+-r_-)/2r_+^2$
and, once again,  $Z^3, Z^4, Z^5$ are three-dimensional spherical coordinates as in 
equation (\ref{SGEMS}). The $Z^6$ direction is timelike as in the AdS-Schwarzschild case. 
We note that this embedding
is not valid for an extremal black hole but we can deduce the temperature in 
the extremal case by taking the limit $r_+\to r_-$ in the final answer.

The result for the square of the free-fall temperature is
\begin{equation}
T_{FFAR}^2 = \frac{1}{16\pi^2r_+^2}\frac{(1-b)^2(1+x+x^2+x^3) - 4bx^4 + 4b^2x^5}{1-bx},
\label{RNtempsq}
\end{equation}
where $x=r_+/r$ and $b=r_-/r_+$. Asymptotically far from the black hole the free-fall 
temperature reduces to the Hawking temperature,
\begin{equation}
\lim_{x\to 0} T_{FFAR} = \frac{r_+-r_-}{4\pi r_+^2} = T_H.
\end{equation}
In Figure~\ref{fig:RNslice} the square of $T_{FFAR}$ is plotted for three different values of $b$. 
\begin{center}
\FIGURE{
\includegraphics{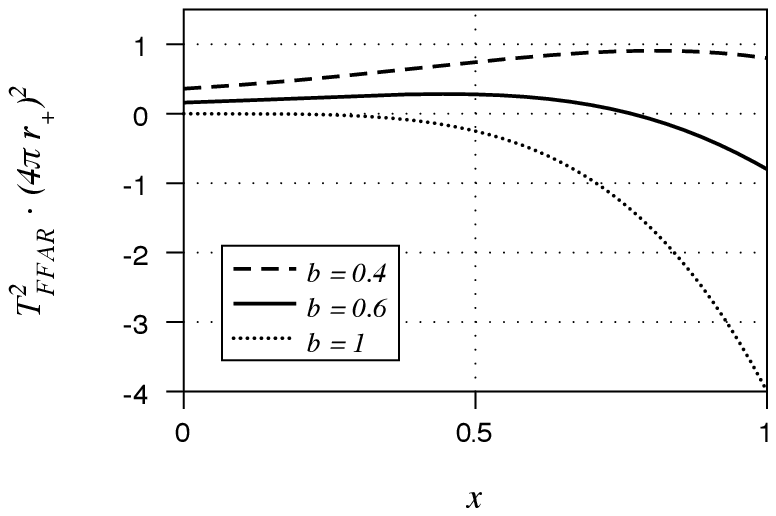}
\caption{$T_{FFAR}^2$ for a Reissner-Nordstr\"om black hole plotted for three different values of 
$b=r_-/r_+$. Geometries with $b\leq 0.5$ have positive temperature everywhere outside the black 
hole but for $b > 0.5$ there is no thermal radiation in a region surrounding the horizon, where 
$T_{FFAR}$ is 
purely imaginary.}
\label{fig:RNslice}
}
\end{center}

In the limit $b\to 0$ the above result for the free-fall temperature reduces to that for the 
Schwarzschild case, while in the extremal limit $r_\pm \to m$ we find
\begin{equation}
T_{FFAR}^2\big\vert_{extremal} = -\frac{m^2}{4\pi^2 r^4}.
\end{equation}
As before, a purely imaginary temperature is interpreted as no thermal radiation and this is 
exactly what is expected for an extremal black hole. Furthermore,
$T_{FFAR}^2\big\vert_{extremal} \to 0$ as $r\to\infty$ 
in agreement with the extremal value of the Hawking
temperature, $T_H=0$.

The curve in the $x$-$b$ plane where $T_{FFAR}^2$ changes sign is given by
\begin{equation}
x_0 =\frac{1-b + \sqrt{1+6b-7b^2}}{4b}, \qquad 0< x_0 \leq 1.
\end{equation}
By analyzing this relation one finds that for $0\leq b \leq \frac{1}{2}$ there is everywhere 
positive temperature outside the black hole but when $\frac{1}{2} < b < 1$ the free-fall
temperature has a single zero at $r=r_0 > r_+$.
Outside $r=r_0$ there is positive temperature but for $r_+<r<r_0$ the free-fall temperature 
is purely imaginary, suggesting that there is no thermal radiation in the region close to the 
horizon of a black hole with high charge-to-mass ratio. This is perhaps not surprising since 
in the extremal limit the near-horizon geometry approaches AdS$_2\times$S$^2$ and a 
freely falling observer in AdS spacetime does not see thermal radiation.

\acknowledgments

This work was supported in part by grants from the Icelandic Research Fund, the University
of Iceland Research Fund, and the Eimskip Research Fund at the University of Iceland. 
E.J.B. wishes to thank NORDITA in Stockholm for hospitality during the completion of this
work.


\begin{thebibliography}{99}

\bibitem{Hawking:1974sw}
  S.~W.~Hawking,
  ``Particle Creation By Black Holes,''
  Commun.\ Math.\ Phys.\  {\bf 43}, 199 (1975)
  [Erratum-ibid.\  {\bf 46}, 206 (1976)].

\bibitem{Unruh:1976db}
  W.~G.~Unruh,
  ``Notes on black hole evaporation,''
  Phys.\ Rev.\  D {\bf 14}, 870 (1976).

\bibitem{Wald:1999vt}
  R.~M.~Wald,
  ``The thermodynamics of black holes,''
  Living Rev.\ Rel.\  {\bf 4}, 6 (2001)
  [arXiv:gr-qc/9912119].

\bibitem{DLads}
  S. Deser and O. Levin,
  ``Accelerated detectors and temperature in (anti-) de Sitter spaces''
  Class.\ Quant.\ Grav. {\bf 14}, L163-L168 (1997)
  [arXiv.org:gr-qc/9706018].    
  
\bibitem{Deser:1998xb}
  S.~Deser and O.~Levin,
  ``Mapping Hawking into Unruh thermal properties,''
  Phys.\ Rev.\  D {\bf 59}, 064004 (1999)
  [arXiv:hep-th/9809159].
  
\bibitem{Santos:2004ws}
  N.~L.~Santos, O.~J.~C.~Dias and J.~P.~S.~Lemos,
  ``Global embedding of D-dimensional black holes with a cosmological  constant
  in Minkowskian spacetimes: Matching between Hawking  temperature and Unruh
  temperature,''
  Phys.\ Rev.\  D {\bf 70}, 124033 (2004)
  [arXiv:hep-th/0412076].  
  
\bibitem{Korsbakken:2004bv}
  J.~I.~Korsbakken and J.~M.~Leinaas,
  ``Fulling-Unruh effect in general stationary accelerated frames,''
  Phys.\ Rev.\  D {\bf 70}, 084016 (2004)
  [arXiv:hep-th/0406080].
  
\bibitem{HL} 
  S. ~Hemming and L. ~Thorlacius, 
  ``Thermodynamics of Large AdS Black Holes,''
  JHEP {\bf 0711}, 086 (2007)
  [arXiv.org:0709.3738].
  
\bibitem{Chen:2004qw}
  H.~Z.~Chen, Y.~Tian, Y.~H.~Gao and X.~C.~Song,
  ``The GEMS Approach to Stationary Motions in the Spherically Symmetric
  Spacetimes,''
  JHEP {\bf 0410}, 011 (2004)
  [arXiv:gr-qc/0409107].

\bibitem{Goenner}
  H. F. Goenner, 
  ``Local Isometric Embedding of Riemannian Manifolds and Einstein's Theory of Gravitation.'' 
  \emph{General Relativity and Gravitation,} 
  edited by A. Held, Plenum, New York, (1980) 441.  
  
\bibitem{analysis}
  Y. Choquet-Bruhat, C. DeWitt-Morette and M. Dillard-Bleick,
  \emph{Analysis, manifolds and physics,}
  North-Holland, Amsterdam (1987).
  
\bibitem{Fronsdal:1959uu}
  C.~Fronsdal,
  ``Completion and Embedding of the Schwarzschild Solution,''
  Phys.\ Rev.\  {\bf 116}, 778 (1959).
  
\bibitem{Unruh:1977ga}
  W.~G.~Unruh,
  ``Origin Of The Particles In Black Hole Evaporation,''
  Phys.\ Rev.\  D {\bf 15}, 365 (1977).

\bibitem{Witten:1998zw}
  E.~Witten,
  ``Anti-de Sitter space, thermal phase transition, and confinement in  gauge
  theories,''
  Adv.\ Theor.\ Math.\ Phys.\  {\bf 2}, 505 (1998)
  [arXiv:hep-th/9803131].

\end{thebibliography}
\end{document}